\documentclass[manuscript]{aastex}
\usepackage{amssymb,amsmath, morefloats, natbib, graphicx}

\newcommand{\eg}{{\it e.g.}}
\newcommand{\ie}{{\it i.e.}}

\newcommand{\etc}{{\it etc}}
\newcommand{\etal}{{\it et al.}}

\newcommand{\PS}{\protect \hbox {Pan-STARRS}}
\newcommand{\PSone}{\protect \hbox {Pan-STARRS1}}

\shorttitle{Asteroid Trail Fitting}
\shortauthors{Veres\v{s} \etal}

\begin{document}


\title{Improved Asteroid Astrometry and Photometry \\with Trail Fitting}


\author{Peter Vere\v{s}\altaffilmark{1}, Robert
  Jedicke\altaffilmark{1}, Larry Denneau\altaffilmark{1} and Richard
Wainscoat\altaffilmark{1}}
\affil{Institute for Astronomy, University of Hawaii at Manoa,
  Honolulu, 96822 HI}

\and

\author{Matthew J. Holman\altaffilmark{2}} 
\affil{Harvard-Smithsonian Center for Astrophysics, Harvard
  University, 60 Garden Street, Cambridge, MA 02138}
  
  \and
  
\author{Hsing-Wen Lin\altaffilmark{3}}
\affil{Institute of Astronomy, National Central University, 300 Jhunda Road, Jhungli, Taiwan 32001}
\email{veres@ifa.hawaii.edu}




\begin{abstract}

Asteroid detections in astronomical images may appear as trails due to
a combination of their apparent rate of motion and exposure duration.
Nearby asteroids in particular typically have high apparent rates of
motion and acceleration.  Their recovery, especially on their
discovery apparition, depends upon obtaining good astrometry from the
trailed detections.  We present an analytic function describing a
trailed detection under the assumption of a Gaussian point spread
function (PSF) and constant rate of motion.  We have fit the function
to both synthetic and real trailed asteroid detections from the
\PSone\ survey telescope to obtain accurate astrometry and photometry.
For short trails our trailing function yields the same astrometric and
photometry accuracy as a functionally simpler 2-d Gaussian but the
latter underestimates the length of the trail --- a parameter that can
be important for measuring the object's rate of motion and assessing
its cometary activity.  For trails longer than about 10 pixels ($\ga
3\times$PSF) our trail fitting provides $\sim 3\times$ better
astrometric accuracy and up to 2 magnitudes improvement in the
photometry.  The trail fitting algorithm can be implemented at the
source detection level for {\it all} detections to provide trail
length and position angle that can be used to reduce the false
tracklet rate.

\end{abstract}





\section{Introduction}

Since the advent of photographic plates asteroids have left their
distinctive trails on astronomical images.  These `vermin of the sky'
leave trailed detections because of the long time durations of the
astronomical exposures and/or the asteroid's fast apparent rate of
motion.  An asteroid's trailing can be eliminated using non-sidereal
tracking at the object's apparent rate of motion but this process
simply transfers the trailing to the field stars while preserving the
telescope system's point spread function (PSF) shape for the asteroid
detection.  Either way, the trailing spreads the total flux from the
object over a larger area than the PSF, causing a reduction in the per
unit area apparent magnitude and signal-to-noise, thereby inducing a
drop in the system's limiting magnitude for fast moving asteroids
\citep[\eg,][]{Kru04,Jed97}.  Thus, dealing with trails on
astronomical images is an unavoidable burden for astronomers who study
the smaller objects in our solar system.

In this work we introduce an analytic function describing a trail due
to an object with an instantaneous Gaussian PSF moving at a constant
rate of motion.  We study the utility and efficiency of fitting the
function to trailed detections using a large sample of both synthetic
and real asteroids.  A trail's morphological parameters as provided by
the trail fitting can be used to significantly reduce the false
tracklet detection rate which means that its use can deliver more
discoveries of fast moving near-Earth objects (NEO), increase the
upper rate of motion limit in automated surveys, and provide improved
astrometry and photometry.

Methods for identifying and characterizing trailed detections were
introduced with the first CCD asteroid survey that used the second
order moments of a 2-d Gaussian to fit the trail
\citep[Spacewatch,][]{Rab91}.  Lately, the characterization of trail
length, astrometry and photometry has been used for artificial
satellites and space debris studies \citep[\eg][]{Kou08,Laas09}. This
work deals specifically with asteroid trail fitting and
characterization.

Although more realistic Lorentz and Moffat PSF profiles
\citep{Kou08,Laas09} have been used for the trail cores the analytical
form of the trail equation has never been published.  It is arguable
whether a more complicated core is necessary given the prevalence of
PSF changes during the exposure time combined with guiding and
tracking irregularities.  To reduce the computational load some
earlier algorithms used a fixed or constrained trail length,
orientation and/or brightness when fitting trails to known objects.
Nowadays, modern all-sky surveys require robust image processing
capable of reliable identification and characterization of transient
detections including asteroid trails.

The Panoramic Survey Telescope and Rapid Response System's prototype
telescope \citep[\PSone;][]{Kai10}), the first of the next generation
all-sky surveys, has submitted almost 3 million asteroid detections to
the Minor Planet Center.  Virtually all the reported astrometry and
photometry are the result of a 2-d Gaussian fit to the
detections using predicted (and fixed) PSF widths based on
extrapolations from nearby field stars.  While the resulting positions
and flux are quite good for slow moving asteroids with PSF-like
detection profiles we will show that the results can be improved
dramatically with a true fit to the trails of fast moving asteroids
like the NEOs.

The trailing loss problem is illustrated in
Fig.~\ref{fig.trailingLoss} which shows how the error in the measured
or reported magnitude drops as a function of the trail aspect (the
ratio of the trail's length to its width).  The figure clearly shows
that the trailing loss can induce errors in the flux measurement that
are much larger than the statistical uncertainty even for `stubby'
trails - those that are trailed only a few times more than their
widths.  A 2-d symmetric Gaussian fit to the trail is clearly bad.  In
theory, an aperture magnitude should work very well as long as the
aperture is rectangular with its long axis aligned with the trail, of
sufficient length to cover the entire trail, and with a width several
times the PSF width.  The catch-22 is that defining the aperture
correctly implies knowing the correct dimension and orientation of the
trail in the first place.  The aperture magnitudes in
Fig.~\ref{fig.trailingLoss} made a simple assumption that the aperture
has a dimension equal to 3$\times$ the PSF width so it yields good
results to trail aspects of $\sim$3 but degrades quickly thereafter.
Assuming larger apertures increases the likelihood of including
background flux and thereby increasing the error on the derived flux.
The real world \PSone\ telescope's Image Processing Pipeline yields
excellent photometry on point sources \citep{Sch12} but fails on
trailed sources because it does not implement a trailing fit.
Presumably all the contemporary asteroid surveys suffer from this
trailing loss at some level.

Due to the worldwide interest in the NEOs, and their relatively high
rate of motion that makes them difficult to recover if too long a time
passes, NEO {\it candidates} are submitted as soon as possible on a
per night basis as a set of $\ge2$ detections known as a `tracklet'
\citep{Kub07}. Subsequent inter-night linking of tracklets of the same
object leads to derivation of the orbit and establishes whether the
candidate is in fact a NEO.

Tracklets are constructed by identifying a set of detections that are
consistent with being the same object that have moved between
exposures.  The false tracklet rate is sensitive to the number of
detections in the tracklet, the false detection rate, the maximum
allowed rate of motion and acceleration, the astrometric and
photometric error on each detection, the availability of the detections'
morphological parameters, \etc.  In practice, the false tracklet rate
is surprisingly high, usually due to systematic (rather than
statistical) false detections.  To reduce the false tracklet rate the
\PSone\ Moving Object Processing System \citep[MOPS;][]{Den07} implements
an upper limit on an object's rate of motion of 4~deg/day for
tracklets containing $\ge3$ detections and only 0.6~deg/day for
tracklets containing just 2 detections.  \PSone\ and contemporary
surveys often rely on human observers to identify and verify trails
and \PSone\ is effectively limited to detecting objects with trails
$<50$~pixels due to the difficulty in identifying trails and the
false tracklet explosion when attempting to link detections of objects
that are moving quickly.   Fitting all the detections to
a trail fitting function helps in every way to reduce the false
tracklet rate.

\section{The trailing functions}
\label{s.TrailFunctions}

\subsection{PSF-convolution trail function}
\label{ss.PSFconvolutionTrailFunction}

An object moving with a constant apparent angular rate of motion
$\omega$ (arcsec/second) in a CCD image with a detector pixel scale
$p$ (arcsec/pixel) in an exposure of time $T$ (seconds) leaves a trail
of length $L = \omega T / p$ pixels.  When $L$ is on the scale of the
PSF width the source detection becomes non-PSF like --- trailed.  We
approximate the trail as the convolution of an axisymmetric Gaussian
PSF of width $\sigma$ moving at a constant rate in a direction $x'$
that is rotated with respect to the image reference frame $(x,y)$. The
flux of the trail at any point in the $(x',y')$ orthogonal coordinate
system is then
\begin{equation}
f_T(x',y') = b(x',y') + \frac{\Phi}{L} \frac{1}{\sqrt{2\pi\sigma^2}}
           \int_{-L/2}^{+L/2} \exp\left[-\frac{1}{2\sigma^2}\left\{(x'-l)^2+(y')^2\right\}\right] dl
\label{eq.gxpyp}
\end{equation}
\noindent where $\Phi$ is the total photometric flux in the trail and
$b(x',y')$ is the background flux at the same point.  The flux in any
pixel $(i,j)$ is strictly the integral of $g(x',y')$ over the bounds
of the pixel but in images that are not under-sampled the function
changes slowly enough across the pixel that we can assume the pixel
flux is simply the flux at the center of the pixel.

This equation can be rewritten using the error function
$\mathrm{erf}(z)=\frac{2}{\sqrt\pi}{\int_0^z e^{-t^2}\,dt}$,
rotating the trail to the image reference frame through an angle
$\theta$ with respect to the $+x$-axis, and translating the trail's
centroid to $(x_0,y_0)$ such that
\begin{equation}
\begin{split}
x'=(x-x_0)\cos\theta-(y-y_0)\sin\theta \\
y'=(x-x_0)\sin\theta+(y-y_0)\cos\theta
\end{split}
\label{eq.rotation}
\end{equation}
 \noindent yielding the trail equation
\begin{equation}
\begin{split}
f_T(x,y)=b(x,y)+\frac{\Phi}{L}\frac{1}{2\sigma\sqrt{2\pi}} \exp\left[-\frac{((x-x_0)\sin\theta+(y-y_0)\cos\theta))^2} {2\sigma^2}\right]  \\
 \left(\mathrm{erf}\left[\frac{(x-x_0)\cos\theta+(y-y_0)\sin\theta+L/2 }{\sigma\sqrt{2}}\right] - \mathrm{erf}\left[\frac{(x-x_0)\cos\theta+(y-y_0)\sin\theta-L/2 }{\sigma\sqrt{2}}\right]\right).
 \end{split}
\label{eq.trail}
\end{equation}

It is well known that an axisymmetric Gaussian profile underestimates
the flux in real PSFs and other more elaborate PSF models such as the
Lorentz or Moffat functional forms could be used (or even empirical
functions based on the realized PSFs within an image
\citep[\eg][]{Tru01}).  Any core PSF function can be convolved with a
line to provide a trail function though most will of necessity need to
be solved numerically.  However, the utility of doing so is
questionable given that in real images the PSF shape and flux are
typically not constant in time --- the object leaving a trail may
itself be changing its intrinsic flux; the seeing can change during
the exposure as may the optical system's transmission function due to
system flexure; wind buffeting, tracking and guiding can change the
apparent path of the object on the CCD or its rate of motion across
the pixels; \etc.

\subsection{2-d Gaussian trail function}
\label{ss.2dGuassianTrailFunction}

Elongated detections have classically been recognized when the major
axis 2$^{nd}$ moment significantly exceeds the 2$^{nd}$ minor axis
moment in a fit to a 2-d Gaussian:
\begin{equation}
f_G(x',y')=b(x,'y')+\frac{\Phi}{2\pi\sigma_x'\sigma_y'}
             \exp\left[-\left( \frac{x'^2}{2\sigma_{x'}^2} 
                            +  \frac{y'^2}{2\sigma_{y'}^2}   \right)\right]
\label{eq.2dGuassPrime}
\end{equation}
\noindent with widths $(\sigma_{x'},\sigma_{y'})$.  Rotating the
function using eq.~\ref{eq.rotation} yields
\begin{equation}
f_G(x,y)=b(x,y)+\frac{\Phi}{2\pi\sigma_x\sigma_y}
             \exp\left[-\left( \frac{(x-x_{0})^2}{2\sigma_{x}^2} 
                            +  \frac{(y-y_{0})^2}{2\sigma_{y'}^2}   \right)\right]
\label{eq.2dGuass}
\end{equation}
Fitting a trail to this functional form provides estimates for the
centroid position, second order moments, background, and total flux
--- but not the trail length.

For the 2-d Gaussian we use the major axis
2$^{nd}$ moment as a proxy for the trail length:
\begin{equation}
L \approx L_g = \sigma^2_{x'}=\int\int\ 2x' \int_{-1/2L}^{1/2L}\frac{\Phi}{L} \frac{1}{2\pi\sigma^2}\exp\left[-\frac{1}{2\sigma^2}\left\{(x'-x_0)^2+y^{'2}\right\}\right] dx_0 \; dx' \; dy'.
\label{eq.sigmaxp2}
\end{equation}
Integrating equation~\eqref{eq.sigmaxp2} and dropping the prime notation
we find that:
\begin{equation}
L_{g}=\sqrt{12 \; (\sigma_{x}^2-\sigma_{y}^2)}
\label{eq.Lg}
\end{equation}

\subsection{Acceleration \& phase angle effects}

Ignoring the topocentric acceleration ($a$) in the trail fit causes an
astrometric error at the mid-time of the exposure, the typical time at
which astrometry is reported, of $a \; T^2/8$ if all the acceleration
is parallel to the direction of motion (along-track).  Taking as an upper limit a good
astrometric error on reported observations of $0.1\arcsec$ than $a \;
T^2 / 8 \la 0.1\arcsec$ or $a \la 1\arcsec/\;T^2$ to guarantee that
the acceleration does not induce significant error in the fit.  For
point of reference, typical exposure times for contemporary asteroid
surveys are on the order of 100~s so that accelerations $\ga
10^{-4}$~arcsec/s$^2$ induce significant astrometric error.

\begin{figure}[!ht]
\epsscale{0.8}
\center
\plotone{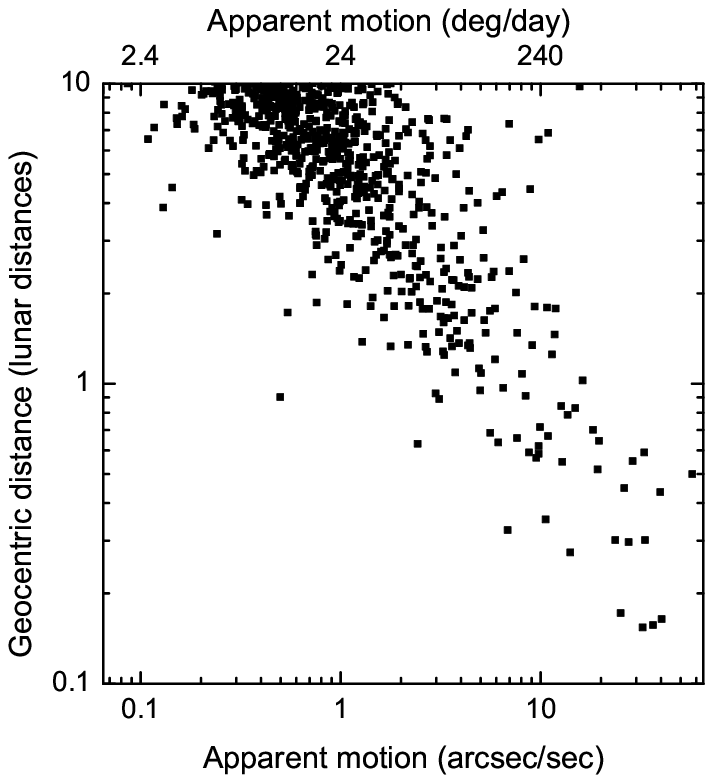}
\caption{Geocentric distance and apparent rate of motion of known
  asteroids at the moment of closest Earth approach between the years
  1900 and 2200 A.D.}
\label{fig.motion_ld}
\end{figure}

\begin{figure}[!ht]
\epsscale{1.0}
\center
\plotone{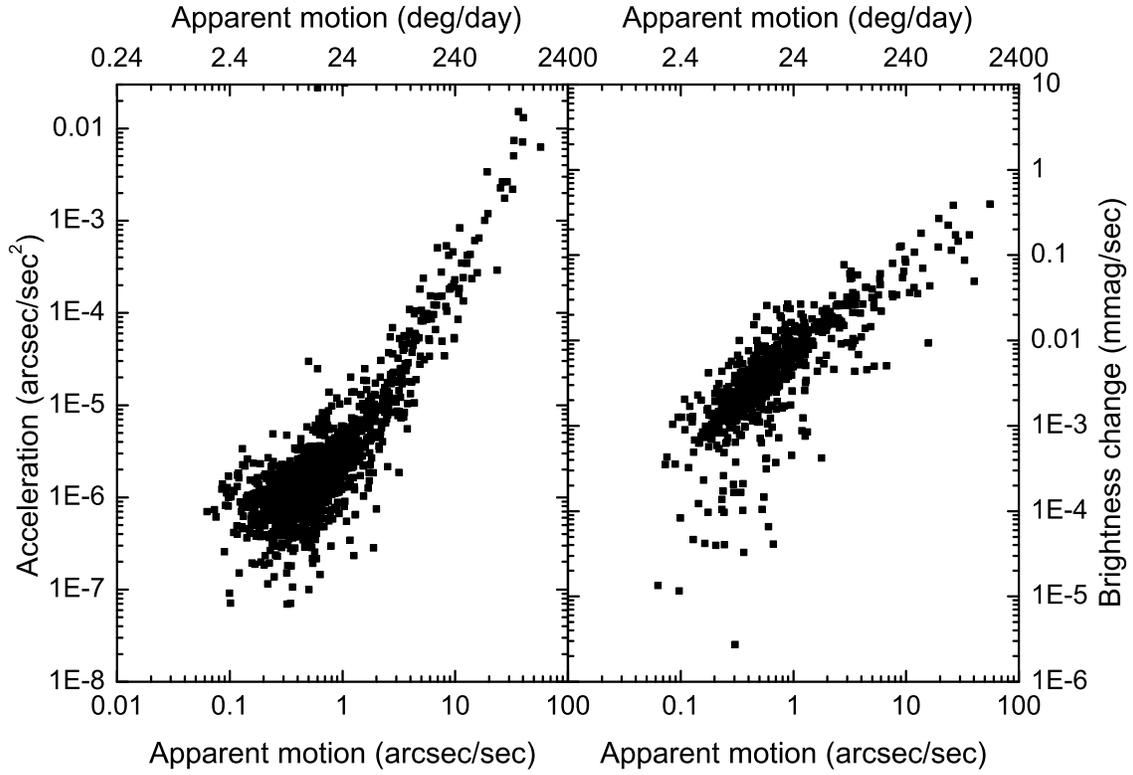}
\caption{Apparent rate of motion, acceleration, and rate of change in
  apparent magnitude of known asteroids at the moment of closest Earth
  approach between the years 1900 and 2200 A.D.  (Left) Apparent rate
  of motion vs. apparent acceleration (Right) Rate of change in
  apparent magnitude vs. apparent rate of motion.}
\label{fig.accel_motion}
\end{figure}

\begin{figure}[!ht]
\epsscale{0.8}
\center
\plotone{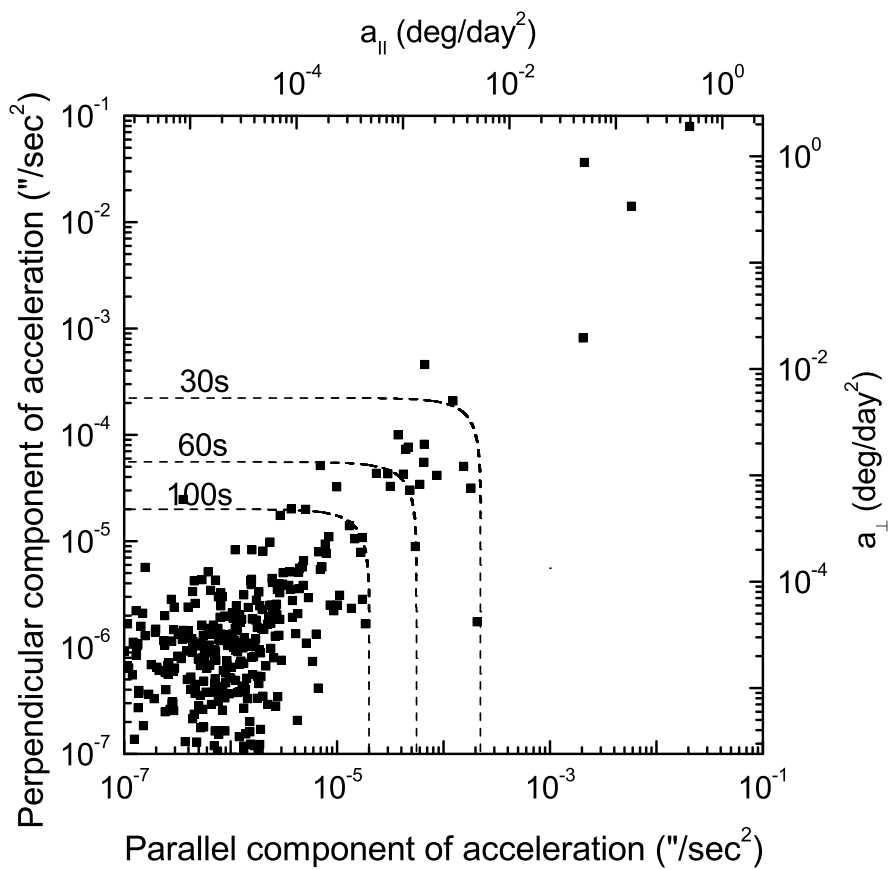}
\caption{Along- and cross-trail components of the
  acceleration of known asteroids at the moment of closest Earth
  approach between the years 1900 and 2200 A.D. Rates of acceleration above and to the right of the dashed curves are large enough to induce $0.1\arcsec$ astrometric error in the fitted trail position at the specified exposure times of 30, 60 and 100 seconds.  \ie\ the technique described in this paper does not induce significant astrometric error for rates of acceleration in the lower left corner of the figure.}
\label{fig.accel_comp}
\end{figure}

\begin{figure}[!ht]
\epsscale{0.8}
\center
\plotone{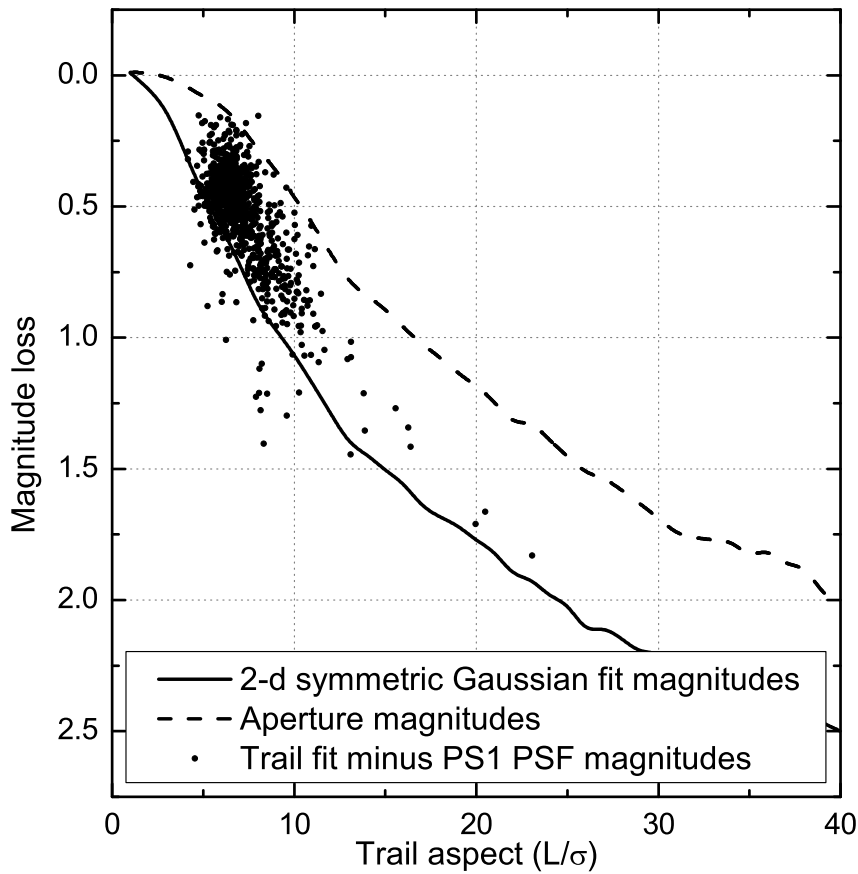}
\caption{Trailing `loss' as a function of the trail's aspect ratio
  (the ratio of its length to width).  A perfect algorithm for
  measuring the magnitude of a trailed detection would provide zero
  magnitude loss.  (solid) The average derived loss in magnitude for
  synthetic trails fit to 2-d symmetric Gaussians with width equal to
  the trail width \ie\ equal to the PSF.  (dashed) The average derived
  loss in magnitude for synthetic trails using square apertures with
  sides equal to 3$\times$ the trail width.  (data points) The
  difference in magnitude between the reported \PSone\ instrumental
  magnitudes and the correct value for the 1,000 longest real trails
  for known asteroids.}
\label{fig.trailingLoss}
\end{figure}

The fastest topocentric accelerations are observed for objects that
are very close to Earth so we examined the apparent rate of motion and
acceleration of known asteroids at the moment of their closest
approach within 10 lunar distances between 1900 A.D. and 2200
A.D. (see figs.~\ref{fig.motion_ld} - \ref{fig.accel_comp}).  Of
the 601 close approaches only about 2\% have accelerations
on the order of the canonical limit of $10^{-4}$~arcsec/s$^2$.
Objects on heliocentric orbits exhibit these accelerations when they
are within about one lunar distance (LD), more typically about 0.5~LDs.  We
note that these objects are moving so quickly across the image plane
that they leave very long trails at typical pixel scales --- since
there is no advantage to long exposures for these trails (the flux per
unit trail length is constant) the solution is simply to use shorter
exposures to eliminate the astrometric impact of neglecting the
objects' acceleration.

Another possible issue for the astrometric fit to a trail is the
effect of changing brightness of the object during the coarse of a
single exposure due to the changing phase angle.
Figure~\ref{fig.accel_motion} shows that the fastest moving objects at
closest approach to Earth change their brightness at $\ga$0.1~mmag/s
so a 100~s exposure would experience $\ga$0.01~mag change in
brightness with a mean of $\sim$0.03~mags.  This effect is much
smaller than typical photometric uncertainty on even slow moving
asteroids and comparable to photometric errors induced by variable
seeing conditions and sky catalogs.  The small asteroids that are
typically observed within 1~LD are also rotating on time scales
comparable to typical exposure times and the rotation amplitudes are
much larger than the phase angle effects \citep{Her11}.  Furthermore,
linear changes in the brightness will probably not induce large errors
in the fitted centroid --- only non-linear changes in the flux that
occur at an even smaller level.  We conclude that the astrometric
impact of changing phase angle is extremely small and, in the worst
cases as described above, can be ameliorated simply by using shorter
exposure times.

Thus, while this work is strictly applicable only to zero acceleration
trails with no change in brightness during the exposure, the vast
majority of objects exhibit negligible acceleration and brightness
change within the exposure time and our technique can be applied
without astrometric impact.  The method is applicable as long as $a \;
T^2 / 8$ is less than the typical astrometric uncertainty which can
almost always be achieved simply by decreasing the exposure time.

\section{Trail fitting performance}
\label{s.TrailFittingPerformance}

We measured the performance of the trail fitting functions on
synthetic and real trails in \PSone\ images.  The synthetic trails were
created to mimic the real \PSone\ trails but allowed us to control the
exact flux, position, length and orientation of the trails to
determine how well the fitting procedure reproduce the generated
values.

We fit the trails to the functions using the IDL procedure {\bf
  mpfit2dfun} \footnote{Markwardt IDL library,
  \url{http://www.physics.wisc.edu/~craigm/idl}} that employes the
Levenberg-Marquardt least-squares fitting technique to minimize the
variance between the trail and the model \citep{Lev44,Mar63}.

We imposed several constraints on the fitting procedure to reduce the
computation time and reduce the likelihood of non-sensical results.
We set the maximum number of iterations to 50 after determining that
all good fits were identified by the algorithm in $<50$ iterations and
imposed limits on the ranges of the fit parameters --- $\sigma$, $L$,
$x_0$, $y_0$.  The trail width $\sigma$ is constrained to be within
the range of stellar PSF widths from all \PSone\ images
(1.5~pix~$<\sigma<$4.5~pix), $L$ must be smaller than the image
dimension, and the centroid position ($x_0$, $y_0$) must be within 20
pixels of the center of the image.  The last constraint was
implemented because the image `source' detected by the source
detection algorithm was placed at the image center.

The resulting fit returned the best values of the 7 fit parameters,
$\bar{x}=(b, \Phi, L, \sigma, \theta, x_0, y_0)$, and their associated
uncertainties, $\delta\bar{x}=(\delta b, \delta \Phi, \delta L, \delta
\sigma, \delta \theta, \delta x_0, \delta y_0)$.  We then calculated
the $\chi^2$ of the fit in a `trail aperture':
\begin{equation}
\chi^2 = \sum_{i,j}^{aperture} \frac{[f_{fit}(i,j;\; \bar{x})-f(i,j)]^2}{\delta f_{fit}(i,j;\bar{x},\delta\bar{x})}
\label{eq.chi2}
\end{equation}
where $f(i,j)$ is the actual flux in the pixel $(i,j)$, $f_{fit}(i,j)$
is the value of the fit function at pixel $(i,j)$ and $\delta
f_{fit}(i,j)$ is the propagated uncertainty in the fit at the same
pixel.  The `trail aperture' over which the summation is performed has
a rectangular shape with center at the fitted centroid position $(x_0,
y_0)$, length of $L+6\sigma$, width of $3\sigma$,
and the long axis of the rectangular aperture is aligned with the
major axis of the fitted trail.  The reduced chi-squared is
$\chi_{red}^2=\chi^2/(N-7-1)$ where the denominator is the number of
degrees of freedom in the fit when ${\it N}$ is the number of the
pixels in the aperture.  We remind the reader that $\chi_{red}^2 \sim
1$ for a good fit while much larger or smaller values indicate a bad
fit.

In the following sections we quantify the trail fitting algorithms'
performance as a function of the trails' statistical significance:
$S/N=\frac{S}{\sqrt{S+B}}$. The signal $S$ is the integrated flux
$\Phi$ of the source as measured by the fit or as generated for
synthetic trails.  The background $B$ is the integrated total
background flux in the trail aperture including readout noise, dark
current, sky, and other sources of flux that are not due to the
trailed object.

\subsection{Generating synthetic trails}
\label{ss.generatingSyntheticTrails}

Synthetic trails were created using eq.~\ref{eq.trail}.  In most
astronomical images the background is dominated by the sky and over a
small region near a trail the sky is generally flat after correcting
for any flat-field issues.  Thus, in the absence of noise the flux at
a given pixel is $f_{syn}(i,j)=f(i,j;\bar{x})+B$ where $B$ is a
constant background level.  In the presence of random statistical
noise the signal at the pixel is
$f'_{syn}(i,j)=f_{syn}(i,j)+\mathrm{ran(1)}\sqrt{f_{syn}(i,j)}$ where
$\mathrm{ran}(1)$ is a random number from the normal distribution with
centroid 0 and width 1.

\begin{figure}[!ht]
\epsscale{1.0}
\plotone{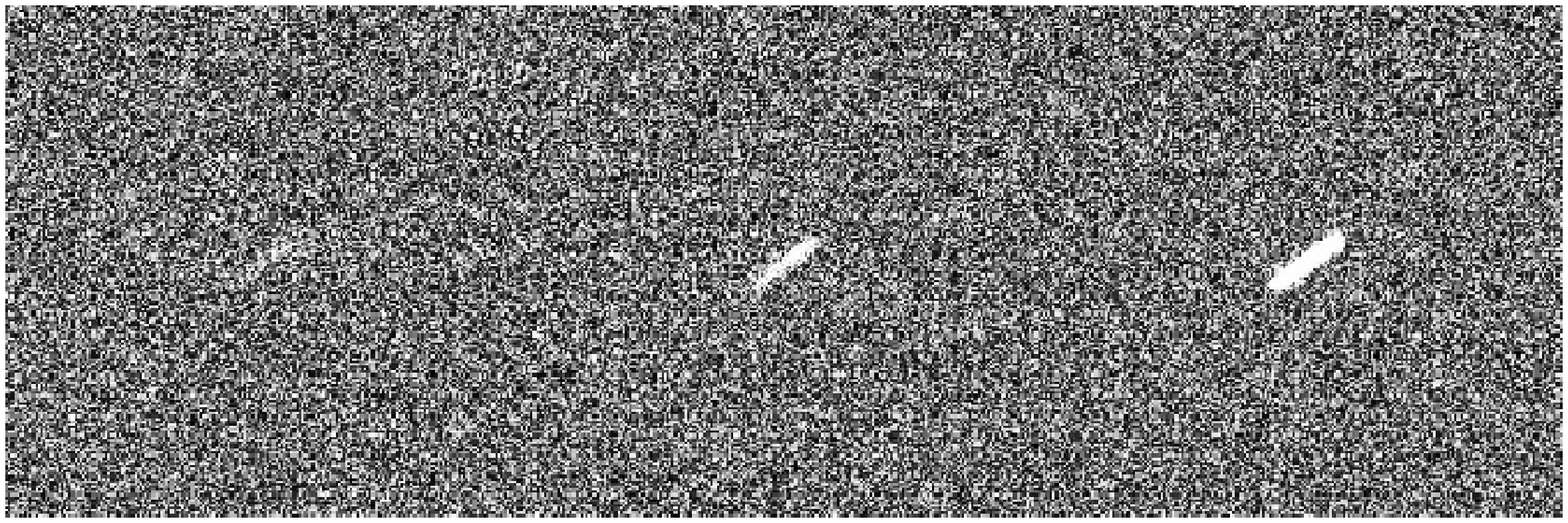}
\caption{Three synthetic images of synthetic trails with $S/N$= 7, 25,
  85.  All the trails have $L=30$~pixels, $\sigma$=2~pixels and
  $\theta=\pi/5=36\deg$. }
\label{fig.trailExamples}
\end{figure}

Figure~\ref{fig.trailExamples} provides a few examples of synthetic
trails where the only difference between them is their $S/N$.  The
faintest trail at $S/N=7$ is only visible as a faint smudge and the eye
is guided to it because we know that the trail lies at the center of
that sub-image.  Detecting these trails with automated software is not
simple --- measuring their properties once detected is the purpose of
this work.

\begin{figure}[!ht]
\epsscale{0.8}
\center
\plotone{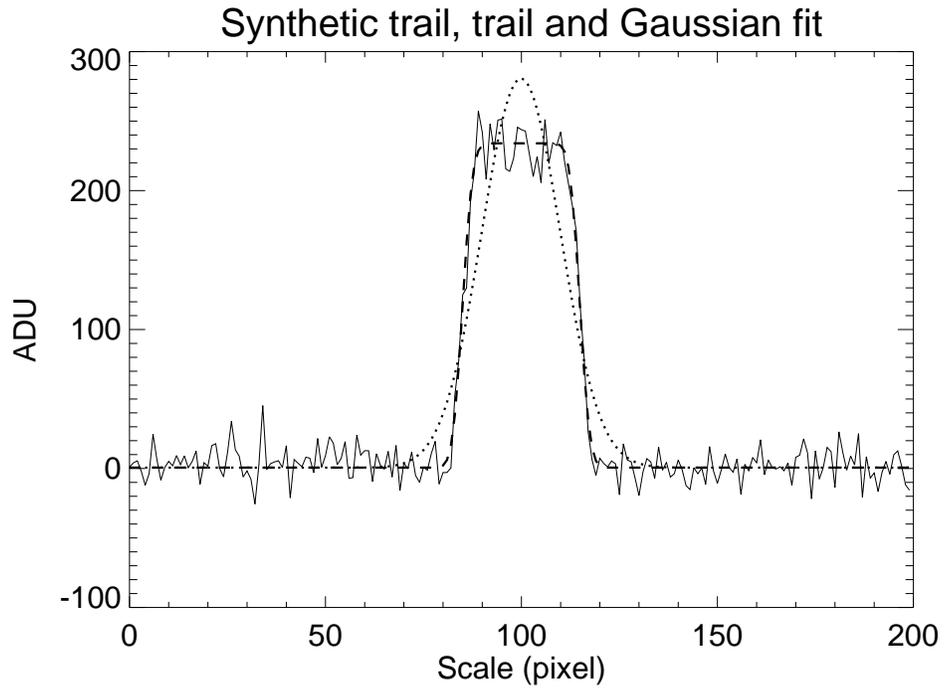}
\caption{Profile of a synthetic trail (L = 30, $\sigma = 2$, S/N=70)
  including noise (solid) along its major axis along with the
  generated trail's shape (dashed) and the result of a 2-d 
  Gaussian fit (dotted).}
\label{fig.trailProfile}
\end{figure}

\begin{figure}[!h]
\epsscale{.50}
\center
\plotone{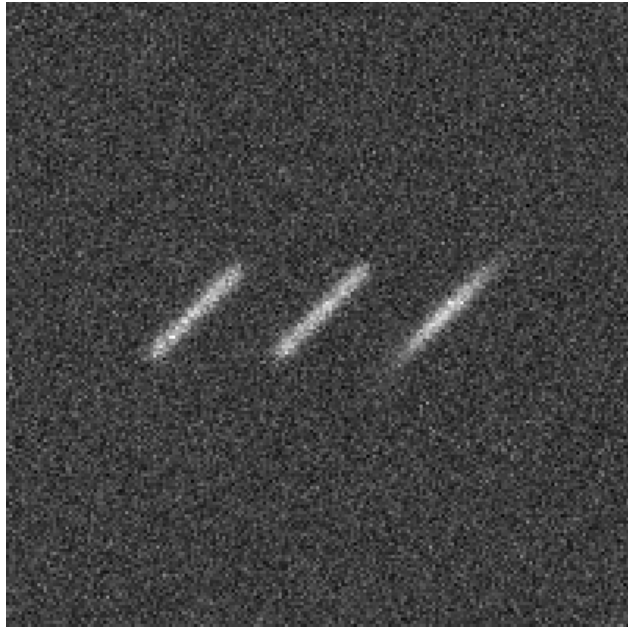}
\caption{(left) Generated synthetic trail (L=40, $\sigma=2.5$,
  $\theta=\pi/4$), the same trail reconstructed (center) by the trail
  fit and the same trail reconstructed with a 2-d Gaussian
  fit (right).}
\label{fig.3trailComparison}
\end{figure}

Figure~\ref{fig.trailProfile} illustrates the difference
between the `true' trail's profile (eq.~\ref{eq.trail}) and a fit to a
2-d Gaussian (eq.~\ref{eq.sigmaxp2}).  It is clear that in
this example the latter can provide reasonable estimates of the
trail's relevant parameters --- its flux, centroid and length (perhaps
after correction with a simple multiplicative length-based factor).
We will show later that the 2-d Gaussian's
utility breaks down for longer trails but its qualitative match to the
generated synthetic trails degrades even sooner as shown in
figure~\ref{fig.3trailComparison}.

\subsection{Synthetic trail fitting}
\label{ss.syntheticTrailFitting}

We measured the trail fitting performance on 30,000 synthetic trails
generated with a flat distribution in length ($10\le L \le 50$),
orientation ($-90\arcdeg < \theta \le +90\arcdeg$), width (1.5~pixels~$\le
\sigma \le$~4.0~pixels), and $3 \le S/N \le 65$. The range in trail
length corresponds to apparent motions of 1.3 to 13.3~deg/day in the
\PSone\ Solar System Survey mode with 45~sec exposure times and a pixel
scale of 0.25~arcsec/pixel.  The centroids of the synthetic trails
were fixed at the center of the small synthetic images.

\begin{figure}[!ht]
\epsscale{1.0}
\center
\plotone{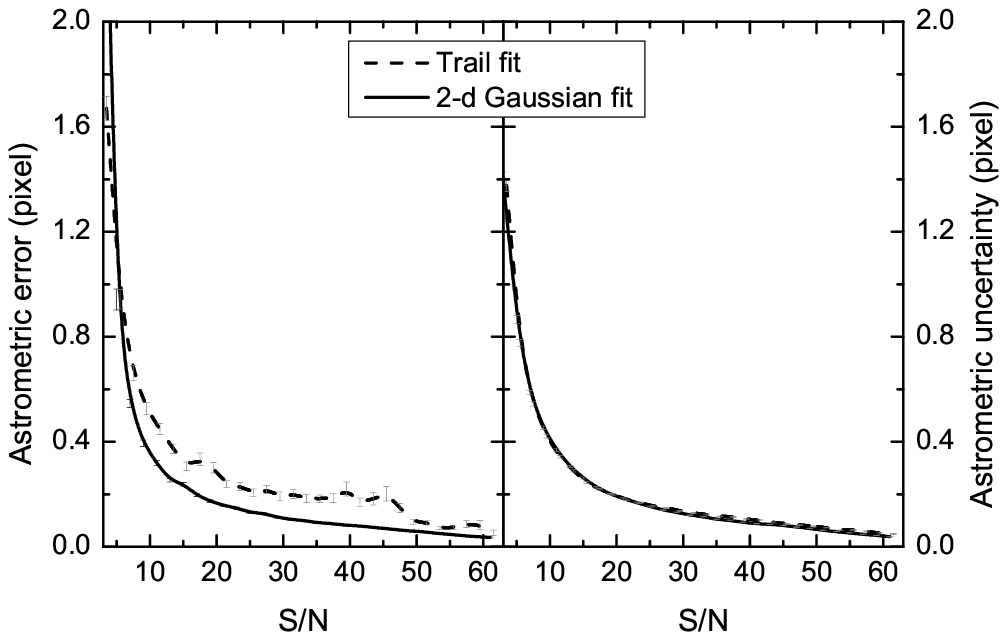}
\caption{(Left) Moving average of the astrometric error for the
    PSF-convolution and 2-d Gaussian trail fits as a
    function of the trail's $S/N$.  The error bars are the standard
    error on the mean.  (Right) Same as on the left but for the
    reported uncertainty.}
\label{fig.astrometricError_s2n}
\end{figure}

Figure \ref{fig.astrometricError_s2n} shows that, as expected, both
astrometric error ($\Delta_x$) and uncertainty ($\sigma_x$) become
smaller as $S/N$ increases and $\sigma_x \ga \Delta_x$. At marginal
$S/N$ the error is on the order of 2 pixels.  With this metric and for
these synthetic trails the 2-d Gaussian fit yields
astrometric errors that range from about $\sim 1.5\times$ to
$\sim2.8\times$ higher than the trail fitted errors for $10<S/N<50$.
While the astrometric errors are much improved with the trail fitting
the reported uncertainties are essentially identical.  Thus, the 2-d
 Gaussian fitting misrepresents the actual error.

\begin{figure}[!ht]
\epsscale{1.0}
\center
\plotone{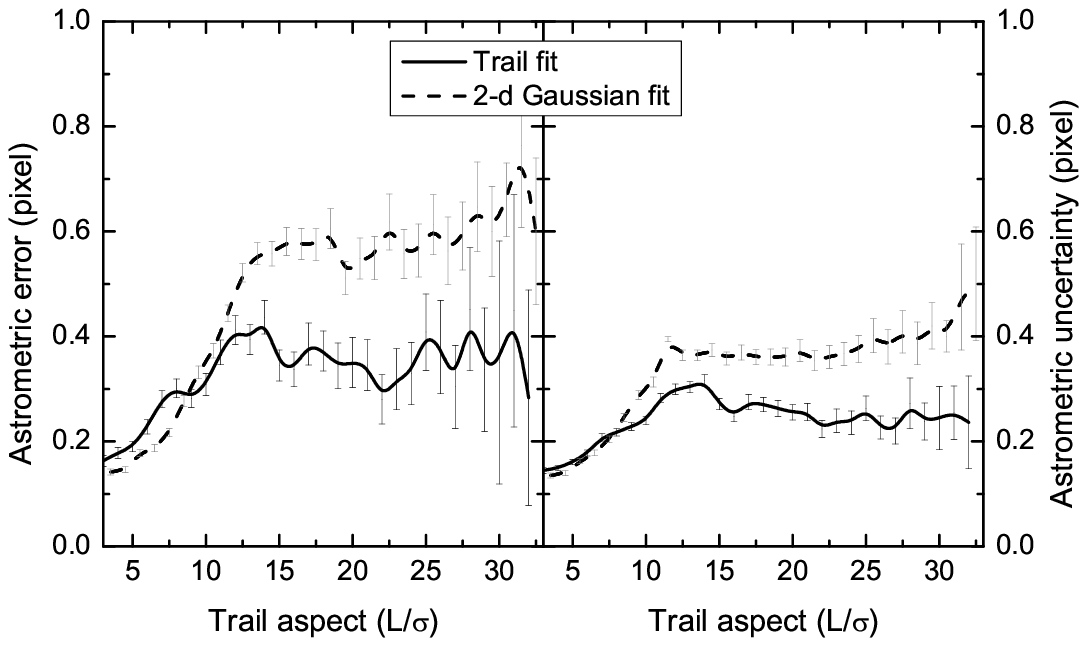}
\caption{(Left) Moving average of the astrometric error for the
  PSF-convolution and 2-d Gaussian trail fits as a function
  of trail aspect ($L/\sigma$) for synthetic trails with $3<S/N<65$.
  The error bars are the standard error on the mean.  (Right) Same as
  on the left but for the reported uncertainty.}
\label{fig.astrometricError_L}
\end{figure}

The utility of the trail fitting function is evident in
fig.~\ref{fig.astrometricError_L} which illustrates how the
astrometric error and uncertainty evolve with the trail `aspect':
$L/\sigma$, \ie\ the length of the trail relative to the PSF.  A
detection with $L / \sigma \sim 1$ appears untrailed while detections
with $L / \sigma \ga 5$ are clearly trailed to eye.  The figure shows
that as a trail's aspect increases both the astrometric error and
uncertainty increase until they plateau for $L / \sigma \ga 10$.  The
plateau values are $\sim$50-100\% higher for the 2-d 
Gaussian fit compared to the trail fitting function.

A problem with trailed asteroid detections is the `loss' of $S/N$ as
the object becomes trailed.  There are two ways to think of this
effect: 1) the peak signal per pixel decreases as $1/L$ while the per
pixel noise remains roughly constant or 2) the total signal in the
trail remains constant as the noise increases because there are more
pixels `under' the trail.  The former interpretation is a trailing
loss because it affects the ability to detect the trail using
peak-pixel detection algorithms.  The latter interpretation affects
the ability of more sophisticated algorithms that attempt to identify
trails by integrating the flux along lines.  In each case the ability
to detect a trailed asteroid decreases due to trailing loss.  The
amount of trailing loss is not the same in the two classes of
algorithms and the implications are too complex to consider in this
work that only discusses fitting the trails once they have been
identified.

\begin{figure}[!ht]
\epsscale{0.80}
\center
\plotone{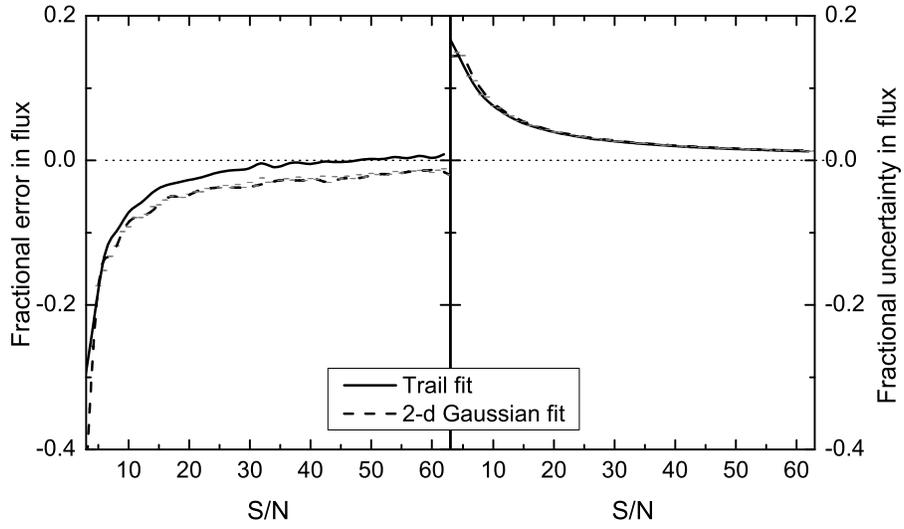}
\caption{(Left) Fractional error and (right) uncertainty of the flux
  derived with the trail and 2-d Gaussian fitting as a function of the
  trail's total $S/N$.  The results include trails of all lengths at
  the same $S/N$ as described in the text.}
\label{fig.s2n_F}
\end{figure}

Once the trail has been detected it is important to correctly measure
the trail's flux and fig.~\ref{fig.s2n_F} illustrates the results of
the flux determination using the trail and 2-d Gaussian fitting
methods.  It is clear that the trail fitting is superior to the 2-d
Gaussian fit but the figure under-represents the improvements with the
trail fitting because it includes trails of all lengths at each {\it
  total} $S/N$.  At small total $S/N$ for longer trails the flux is
spread over many pixels so that the per pixel $S/N$ is much smaller
(the `trailing losses' are described in more detail later in this
section).  In this situation, by far the most common since the number
of trails increases exponentially with decreasing $S/N$, the trail
fitting algorithm dramatically outperforms the 2-d Gaussian method.

\begin{figure}[!ht]
\epsscale{0.75}
\center
\plotone{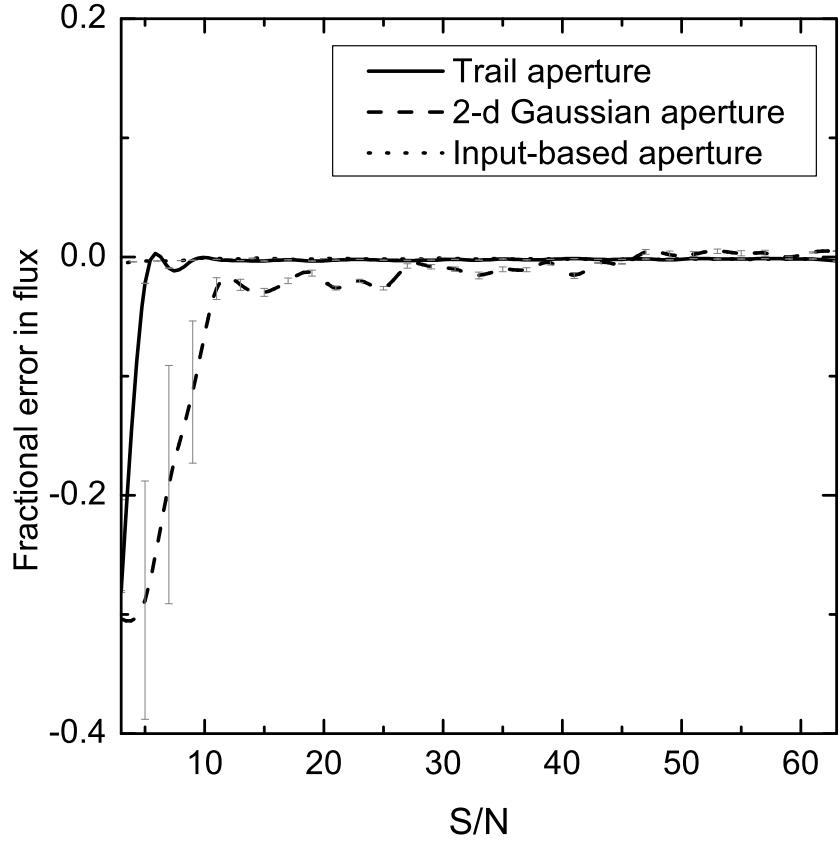}
\caption{Fractional error of the measured trail flux in rectangular
  apertures with size and orientation determined by the two types of
  trail fitting and using the generated (input) parameters versus
  total $S/N$.  The results include trails of all lengths at the same
  $S/N$.}
\label{fig.s2n_aperture}
\end{figure}

Figure~\ref{fig.s2n_aperture} provides a comparison to the aperture
measured photometry using apertures specific to the trail and fit
type.  \ie\ the trail fit uses a rectangular aperture as described in
\S\ref{s.TrailFittingPerformance} and the 2-d Gaussian aperture uses a
similar rectangular apertue but with $L \rightarrow L_g$ and $\sigma
\rightarrow \sigma_g$.  We see that trail photometry using apertures
based on the generated trail lengths and widths reproduces the
expected flux with zero error as expected.  An aperture based on the
trail-fit length and width performs very well for $S/N \ga 5$ while a
flux measurement based on the 2-d Gaussian fit aperture only has
approximately zero error for $S/N \ga 25$.  Once again, since most
trails have small $S/N$-weighted error metric would clearly show that
the trail-fitting technique offers a major advantage over the 2-d
Gaussian fit.

While the primary purpose of this work was to illustrate the
astrometric and photometric improvements when using our
PSF-convolution trail fitting the remaining two parameters, the trail
length and orientation, are important in linking individual detections
on the same night together into `tracklets' \citep{Kub07}.  This is a
common combinatoric explosion when asteroid surveys attempt to link
detections of fast moving asteroids --- the search area increases as
the square of the upper limit on the rate of motion thereby squaring
the number of candidate matching detections.  One might think that the
astrometric and photometric requirements would be sufficient to
eliminate false linkages between the detections but, surprisingly,
there are often far more false detections than statistics would
predict leading to a disturbing number of false tracklets.  The
typical solution is to decrease the upper rate limit to keep the false
tracklet rate manageable and to increase the number of required
detections for a valid tracklet.

Another solution to the tracklet formation problem is to use more
information --- larger areas must be searched as the upper limit on
the rate of motion is increased but the faster moving objects will
also leave trails rather than point source detections.  Those trails
will typically be of approximately the same length (assuming constant
exposure times) and will have roughly the same orientations.  Thus, a
good trail fitting algorithm on candidate detections can dramatically
reduce the false tracklet formation rate.

\begin{figure}[!ht]
\epsscale{1.00}
\center
\plotone{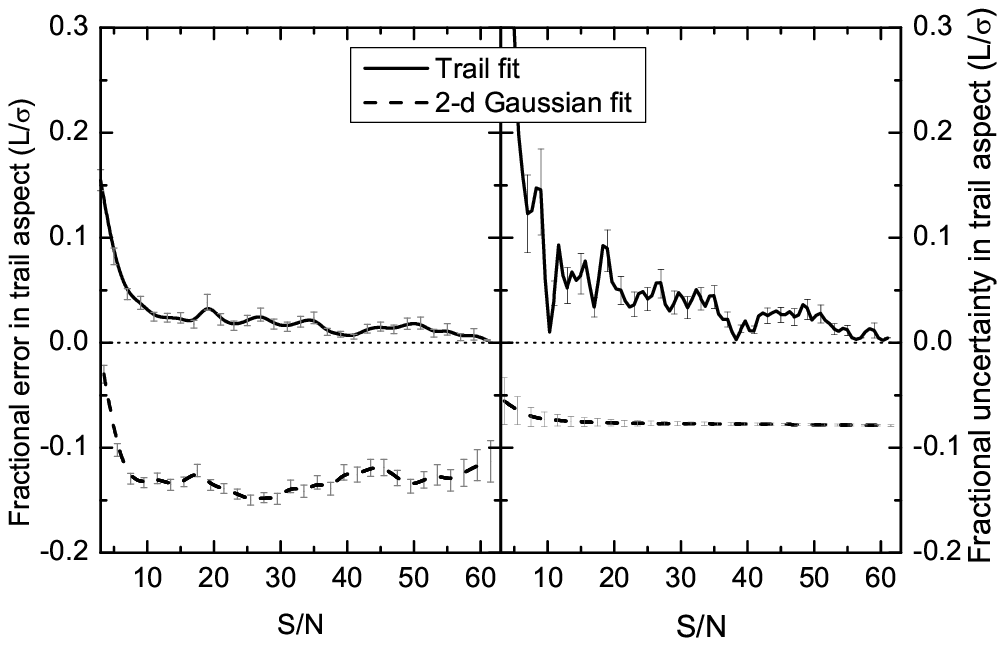}
\caption{(Left) Moving average of the fractional error on the trail
  aspect ($L/\sigma$) for the PSF-convolution and 2-d
  Gaussian trail fits as a function of $S/N$.  The error bars are the
  standard error on the mean.  The trail aspect for these synthetic
  trails ranges from $3 < L/\sigma < 35$.  (Right) Same as on the left
  but for the reported uncertainty.}
\label{fig.s2n_L}
\end{figure}

The ability to correctly measure the length of a trail is illustrated
in fig.~\ref{fig.s2n_L} in terms of the derived trail aspect error and
uncertainty as a function of $S/N$. The length derived with the 2-d
Gaussian is always underestimated by $\ga 10\%$ for $S/N \ga 5$ while
the PSF-convolution trail fitting provides an error of $\la 3$\% in
the same $S/N$ range.  This is not unsurprising as, by its very
nature, the 2-d Gaussian {\it should} always underestimate the trail
length. The trail aspect uncertainty returned by the 2-d Guassian is
always underestimated by about 10\% while the trail fit uncertainty is
always slightly higher than the error as desired.

\begin{figure}[!ht]
\epsscale{1.0}
\center
\plotone{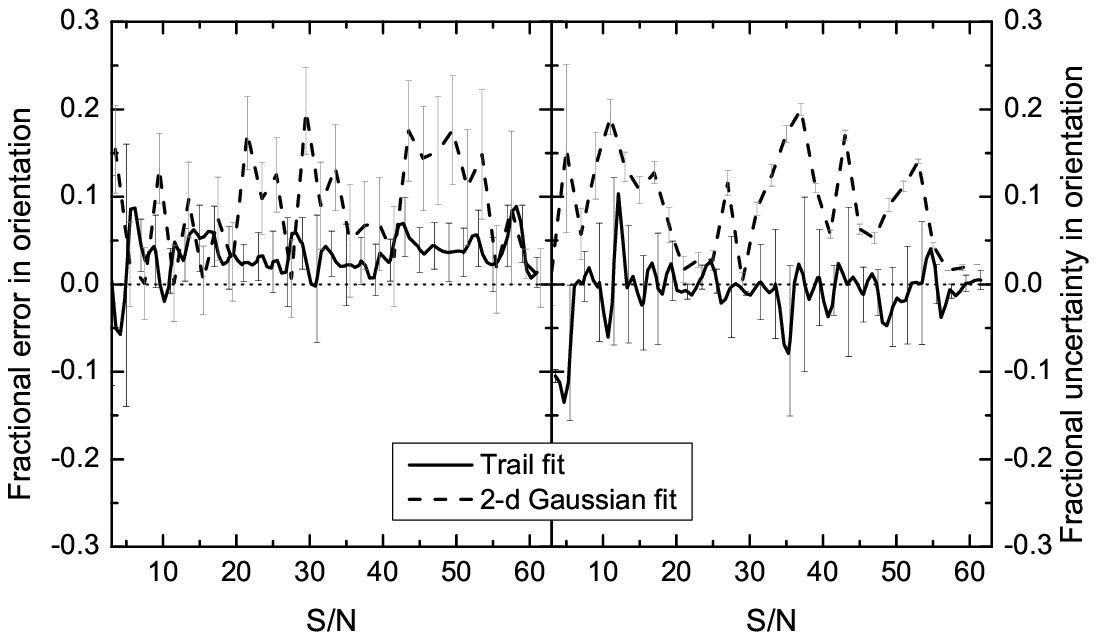}
\caption{(Left) Moving average of the fractional error on the trail
  orientation for the PSF-convolution and 2-d Gaussian
  trail fits as a function of $S/N$.  The error bars are the standard
  error on the mean.  The trail aspect for these synthetic trails
  ranges from $3 < L/\sigma < 35$.  (Right) Same as on the left but
  for the reported uncertainty.}
\label{fig.s2n_angle}
\end{figure}

Similarly, fig.~\ref{fig.s2n_angle} demonstrates that the error in
orientation does not depend on the $S/N$ but that the PSF-convolution
method consistently performs better than the 2-d Guassian
fits.  Since the measured trail lengths/aspects and orientations can
be used to reduce the false tracklet rate it is important to use a
functional form that returns more accurate values and the
PSF-convolution trail fit is clearly superior.

\section{Real trail fitting}
\label{s.realTrails}

We have shown above that our PSF-convolution trail function performs
well on fitting synthetic trails but the true litmus test is how it
performs on real detections in real images with all their systematic
noise problems, dead and bright pixels, edges and gaps, saturated
stars and diffraction spikes, ghost images, cosmic rays, \etc.  

To test our algorithm on real trails we employed asteroids detected by
the \PSone\ telescope and identified by MOPS. The MOPS database contains
hundreds of thousands of known and unknown asteroids of which several
thousand are trailed. MOPS stores 200$\times$200 pixel `postage
stamps' of all \PSone\ detections that were incorporated into tracklets.
The postage stamp FITS file is the difference of two images of the
same portion of the sky exposed at different times \citep{Lup07} and
is therefore already background subtracted.

\begin{figure}[!ht]
\epsscale{.70}
\centering
\plotone{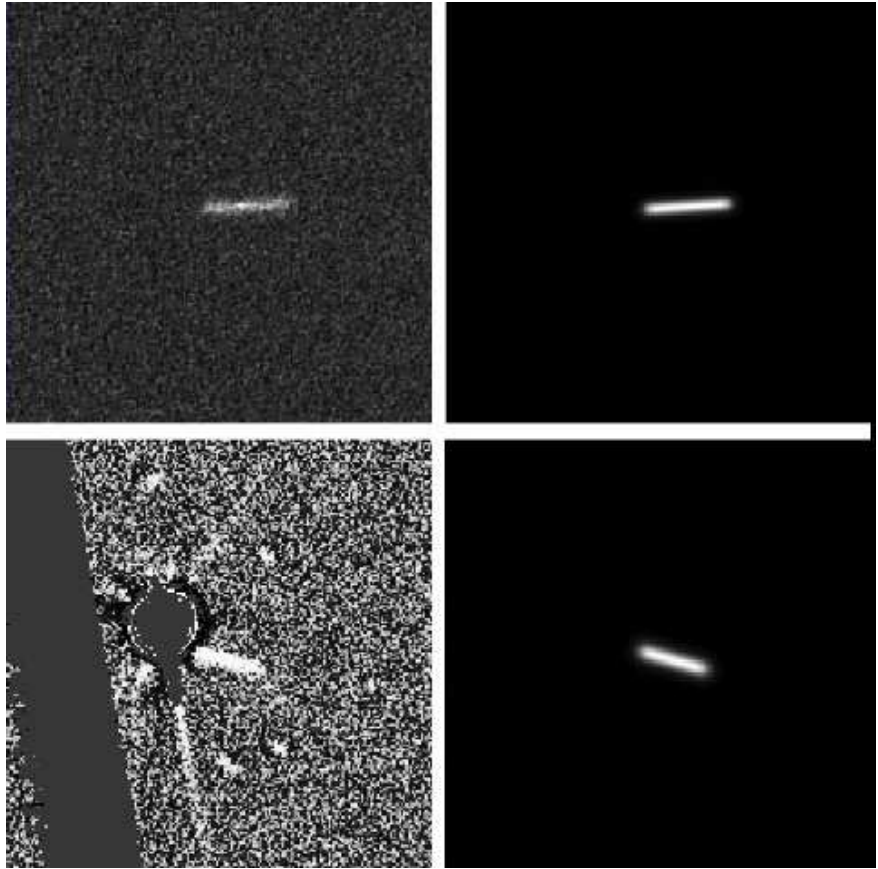}
\caption{(left) Real \PSone\ difference images of trailed asteroids and
  (right) their reconstructed trails from our PSF-convolution trail
  fitting.  (top) An almost ideal case with a nicely trailed detection
  in the center and no obvious image artifacts and (bottom) a more
  typical case where the trailed detection passes through the masked
  pixels of a badly subtracted bright star with diffraction spikes,
  `streaks', and a gap between CCD chips.  Masked or non-existent
  pixels are in grey.}
\label{fig.2realTrails}
\end{figure}

Figure~\ref{fig.2realTrails} provides examples of a couple real
postage stamps --- one ideal and one typical.  In the ideal example
our trail fitting worked perfectly but even for the more typical case
the PSF-convolution trail fit provides a good reconstruction of the
{\it visible} part of the detection.  Even though the trail has been
truncated there is still useful information in its orientation and
length to assist in linkning similar detections on a night into
tracklets.  In cases where the trail is truncated our trail fitting
algorithm raises a flag that there are masked pixels in the aperture.
In these cases the astrometry and photometry are also flagged as
suspect.

To measure the astrometric error for real asteroid trails we selected
the 1,000 longest trails of numbered asteroid detections from the MOPS
database that were submitted to the MPC.  These trails have an average
length of 15 pixels and the longest automatically detected trail in
the sample is 25 pixels long. \eg\ an average trail aspect of $\sim$3.
For comparison, the average length of the 1,000 longest submitted
tracklets irrespective of whether they are known is 18 pixels with the
longest being 43 pixels.  Since the \PSone\ survey has not emphasized
the detection of trailed, fast moving objects, even longer trails
exist in the data that were not detected by the IPP's source detection
algorithms. For instance, we manually identified 114~pixel long
trailed detections of 2012~LZ$_1$ during its close approach to the
Earth.  These and other trails will be identified and measured as the
data are reprocessed but the objects will be impossible to recover.
On the other hand, their astrometry and photometry will be useful for
future precoveries of the objects by \PSone\ and other surveys.

\begin{figure}[!ht]
\epsscale{.70}
\centering
\plotone{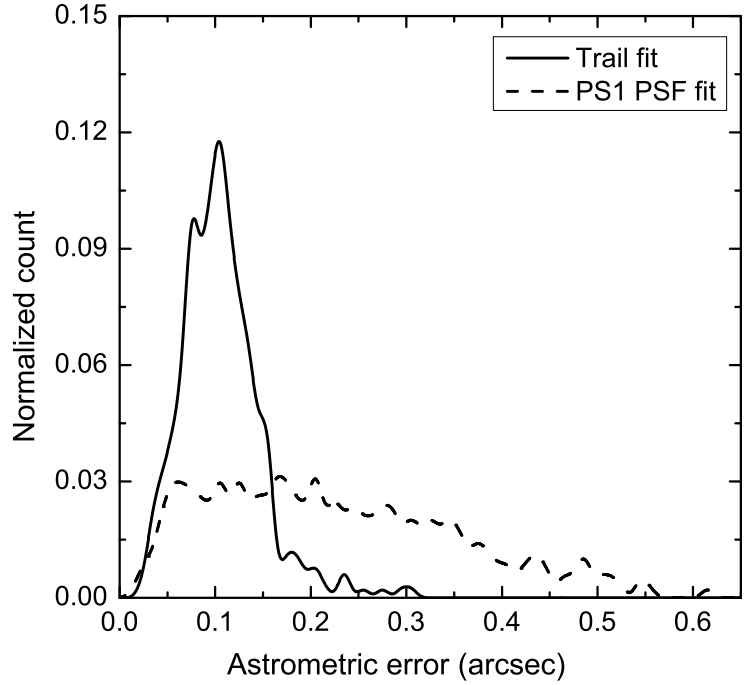}
\caption{(solid) Astrometric error after trail fitting for the 1,000
  longest trailed detections of numbered asteroids that were submitted
  to the MPC by \PSone and (dashed) for the submitted astrometry from
  the \PSone's PSF fitting algorithm.  For comparison, the average
  \PSone\ astrometric error for untrailed detections is
  $\sim0.13$~arcsec \citep{Milani12}. }
\label{fig.astrometricError}
\end{figure}

We calculated the expected position for each of the 1,000 trailed
detections of numbered asteroids at the epoch of observation using
OpenOrb \citep{Gra09} that uses DE406 as input for the perturbing
planets. Since the numbered asteroids have accurate orbits we can use
their calculated positions to measure the astrometric error of our
trail fitting method.  Figure~\ref{fig.astrometricError} shows that
the PSF-convolution trail fitting provides a 2-3$\times$ improvement
in the realized astrometric error over a 2-d Gaussian fit for these
trails.

It is clear that figure~\ref{fig.astrometricError} underestimates the
improvement in astrometric error in the trail fitting because we have
tested the algorithm only on trails for numbered asteroids.  These
objects are on average larger, more distant, and leave shorter trails
(smaller trail aspects) on the image than the smaller, closer,
unnumbered and newly discovered asteroid trails.  Reducing the
astrometric (and photometric) error on newly discovered objects will
have a dramatic impact on the ability to recover the objects shortly
after discovery by 1) decreasing the sky-plane ephemeris uncertainty
and 2) providing better magnitude estimates to improve target
selection by the followup observatories.  \ie\ to better match the
sites' detection capabilities to the target's brightness.

As discussed above, false linkages can be dramatically reduced by
requiring that the detections within a tracklet have the same
orientation (position angle; PA) and roughly the same lengths.  The
fitted length of real trails is not as robust a filter as the PA
because real trails often intersect chip gaps, boundaries, diffraction
spikes, \etc., that tend to truncate the measured trail length.  In
practise, the PA filter constraint must be trail length dependent.

\begin{figure}[!ht]
\epsscale{1.0}
\centering
\plotone{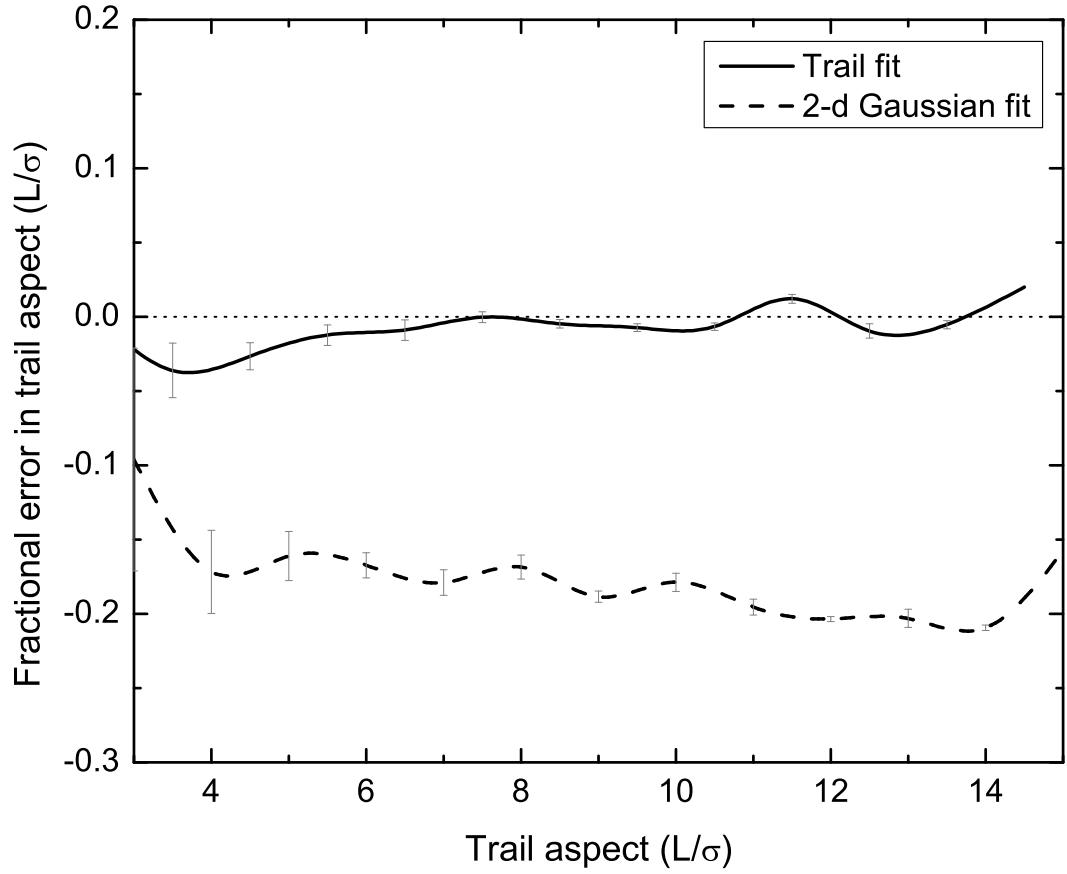}
\caption{Fractional error in the trail aspect $L/\sigma$ for the 1,000
  longest trails for known asteroids for both the (solid) trail
  fitting and (dashed) 2-d Gaussian fits.}
\label{fig.realL}
\end{figure}

\begin{figure}[!ht]
\epsscale{1.0}
\centering
\plotone{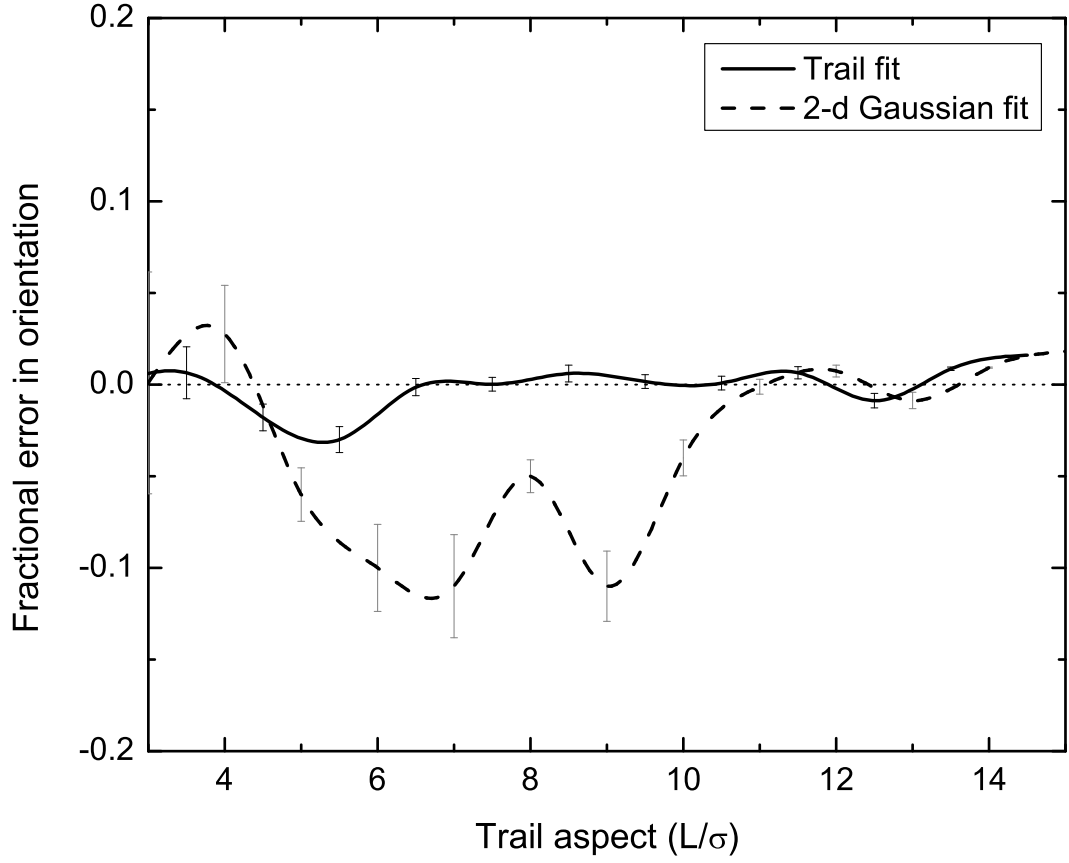}
\caption{Fractional
  error in the trail orientation $\theta$ for the 1,000 longest trails
  for known asteroids for both the (solid) trail fitting and (dashed)
  2-d Gaussian fits.}
\label{fig.real_angle}
\end{figure}

Figures~\ref{fig.realL} and \ref{fig.real_angle} show the relative
error in both measures in our 1,000 trail sample and generally mimic
the results from our studies with synthetic trails as shown in
figs.~\ref{fig.s2n_L} and \ref{fig.s2n_angle}.  The PSF-convolution
trail fit provides lengths (or equivalently, aspects) consistent with
the ephemeris prediction but the 2-d Gaussian derived lengths are
underestimated by $\la 20$\%.  The PSF-convolution trail fit derived
orientation matches the predicted position angle for all aspect ratios
but the 2-d Gaussian-derived orientations show a strange and
unexplained error of up to 10\% for moderate trail aspect ratios.

One of the major problems with trail identification and
characterization is computing time.  Even though the human eye and
mind are tuned to identify these features writing a computer program
to do so in astronomical images and at small $S/N$ has proven to be
challenging.  Fitting our trail function to a source detection that
has already been identified is less time consuming but still $\sim
3\times$ more computationally expensive than simple PSF fitting.  For
example, \PSone\ identifies so many false detections that it is
computationally impossible with the current hardware resources to fit
all identified source detections to the PSF-convolution trail function
even though doing so would dramatically reduce the false detection
rate.

\begin{figure}[!ht]
\epsscale{1.0}
\center
\plotone{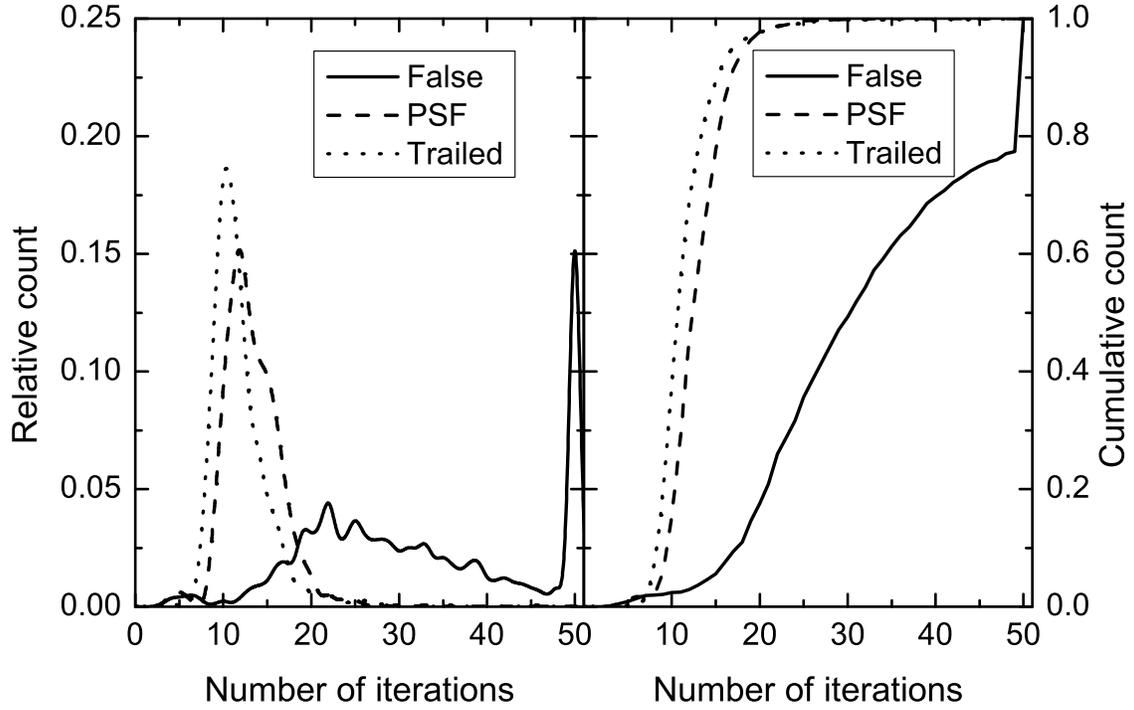}
\caption{Number of iterations needed for fit convergence for trailed,
  PSF-like and false detections.}
\label{fig.niter}
\end{figure}

Instead, MOPS currently links detections based purely on the
astrometry and consistent motion between detections \citep{Milani12} and
only fits the PSF-convolution trailing function to those detections
vetted and selected by a human observer.  Along with the fitted
trail's $\chi^2$ we find that we can select for good trails simply by
limiting the number of iterations in the fit (see
fig.\ref{fig.niter}).  The fits to fully 99\% of real asteroid
detections (both trailed and PSF-like) converged in under 25
iterations while 69\% of false detections do not converge.  (The
number of iterations needed for convergence was tested on a sample of
1,000 trailed, 1,000 PSF-like and 1,000 false detections from \PSone.)
This is a powerful computational savings since the number of false
detections far outnumbers the number of real detections.

\subsection{False detection reduction}

\begin{figure}[!ht]
\epsscale{1.0}
\center
\plotone{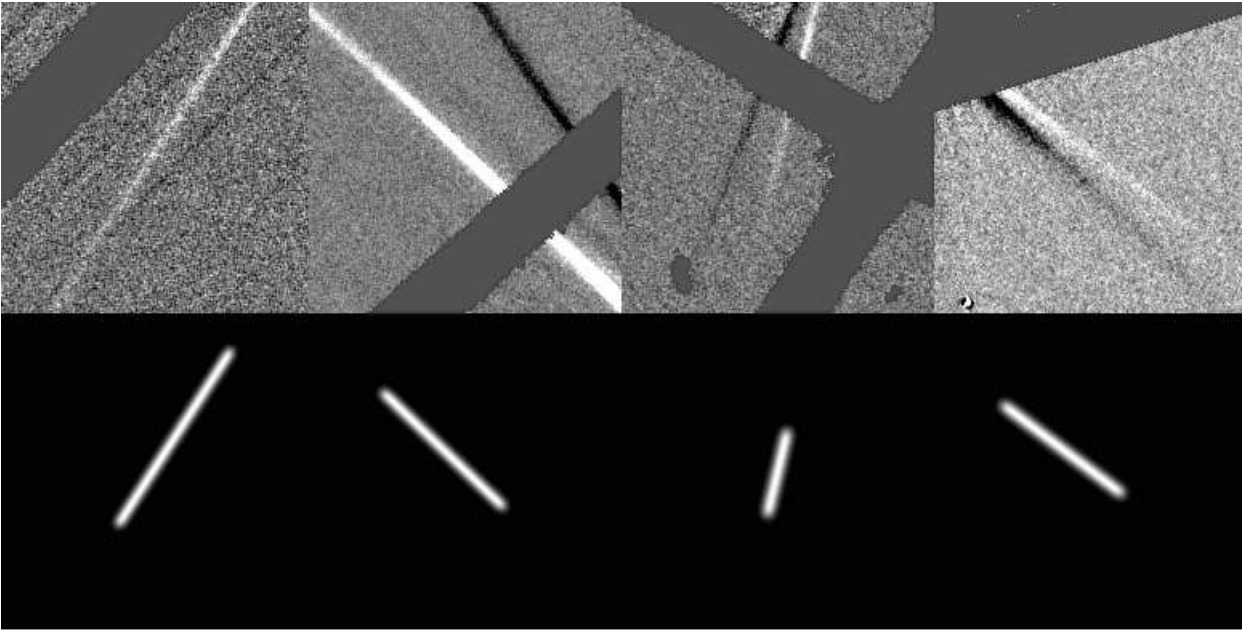}
\caption{Four examples of (top) diffraction spikes in difference
  images and (bottom) their corresponding trail fits.  The difference
  images show both the positive and negative diffraction spikes from
  the same stars but with field rotation between the two images.  The
  trails are only fit to the positive diffraction spikes.  The thick
  gray bands in the difference images represent chip gaps.}
\label{fig.falseDetections}
\end{figure}

Trail fitting is effective at eliminating false detections and even
better at eliminating false tracklets.  In \PSone\ images one of the major
problems is false detections along diffraction spikes as shown in
fig.~\ref{fig.falseDetections}.  Statistical fluctuations along the
length of the diffraction spikes trigger multiple instances of the
PSF-detection algorithm.  These false detections can be eliminated by
subsequently fitting them to our trailing function because the trail
fit returns an unreasonably large RMS and $\chi^2$ and position angles
that are aligned with the expected PA for diffraction spikes in the
image. 

We measured the trail fitting's false detection reduction capability
using postage stamps on $\sim$34,000 detections identified by \PSone's
PSF-detection algorithm.  The maximum number of iterations was set to
25 (as per figure~\ref{fig.niter}) and we required that the fit's
reduced $\chi^2<3$.  These cuts kept $\sim$96\% of real detections and
rejected 98\% of the false detections.

Tracklet rejection was based on cuts suggested by figures
\ref{fig.realL} and \ref{fig.real_angle} \eg\ detections within a
tracklet must have lengths and orientations consistent to within 10\%
and, as discussed above, the fits can not be compromised by contacting
masked detector areas.  We achieved a false tracklet rejection rate of
99\% while the fraction of rejected real tracklets was only $\sim$1\%.
The tracklet rejection rate can be higher than the detection rejection
rate because the same false detections can appear in multiple
tracklets --- each of which is rejected by identifying the errant
detection.

\section{Conclusions}

We provide the analytic form for an asteroid trailing function
assuming that a symmetrical Gaussian PSF is moving at constant speed
during an exposure.  We then demonstrate that this function provides
accurate photometry and astrometry --- much better than the values
derived using 2-d Gaussian fits to the trails and much better than the
values currently being provided by the \PSone\ survey.

The trailing function provides astrometry with $>3\times$ smaller
errors than are typically reported by the surveys.  The photometry
with the trail fitting improves by an even larger factor.  It is clear
that the trail fitting function should be implemented whenever
possible --- even when fitting stationary objects since they are
nothing but trails of length $L=0$.  We note that there are extra
benefits to using the trailing function for all image sources because
the length and orientation can be used to diagnose optical and focus
effects within the image plane.

The trail fitting morphological parameters can eliminate on the order
of 99\% of false tracklets by requiring that detections within the
tracklet have the same lengths, orientations, and fluxes.  The derived
trail lengths and position angles can also help to identify false
detections that are saturated streaks and diffraction spikes.

The trail fitting computation time might be an issue because it can be
significantly longer than for the 2-d Gaussian.  A hybrid method that
will be implemented by \PSone\ is to apply the trail fitting only to
detections with large second order moments and/or excessive flux in
the aperture that might indicate a trail.

\section{Acknowledgements}

We would like to thank Jan Kleyna, David Tholen, Marco Micheli, Henry
Hsieh, and Eugene Magnier from the Institute for Astronomy at the
University of Hawaii for their support and helpful discussions.  We
thank an anonymous reviewer for helpful feedback.  We also thank the
PS1 Builders and PS1 operations staff for construction and operation
of the PS1 system and access to the data products provided. The
\PSone\ Surveys have been made possible through contributions of the
Institute for Astronomy, the University of Hawaii, the \PS Project
Office, the Max-Planck Society and its participating institutes, the
Max Planck Institute for Astronomy, Heidelberg and the Max Planck
Institute for Extraterrestrial Physics, Garching, The Johns Hopkins
University, Durham University, the University of Edinburgh, Queen's
University Belfast, the Harvard-Smithsonian Center for Astrophysics,
and the Las Cumbres Observatory Global Telescope Network,
Incorporated, the National Central University of Taiwan, and the
National Aeronautics and Space Administration under Grant
No. NNX08AR22G issued through the Planetary Science Division of the
NASA Science Mission Directorate.

\bibliographystyle{pasp}
\bibliography{veres}

\clearpage

\clearpage

\clearpage

\clearpage

\clearpage

\clearpage

\clearpage

\clearpage

\clearpage

\clearpage

\clearpage

\clearpage

\clearpage

\clearpage

\clearpage

\clearpage

\clearpage

\clearpage

\clearpage

\end{document}